\documentclass[aps,prx,longbibliography,twocolumn,amsmath,amssymb,floatfix,eqsecnum,superscriptaddress]{revtex4-1}

\usepackage{amsmath}
\usepackage{amssymb}
\usepackage{amsthm}
\usepackage{mathtools}
\usepackage{mathdots}
\usepackage{amsfonts}
\usepackage{bbm}

\usepackage[pdftex]{color} 
\usepackage{graphicx}
\usepackage{dcolumn} 
\usepackage{bm} 
\usepackage{float} 
\usepackage[driverfallback=dvipdfm]{hyperref} 
\usepackage{longtable}
\usepackage{ulem}   
\allowdisplaybreaks

\usepackage{natbib}
\usepackage{graphicx}

\usepackage{amssymb}
\usepackage{mathbbol}
\usepackage{amsmath,amsfonts,amsthm,mathtools} 
\usepackage[makeroom]{cancel}
\usepackage{soul}
\usepackage{bbm}
\usepackage{xcolor}

\usepackage{subfigure}

\usepackage{hyperref}

\renewcommand\[{\begin{equation}}
\renewcommand\]{\end{equation}}

\begin{document}

\title{Chiral Superconductivity in UTe\(_2\) via Emergent \(C_4\) Symmetry and Spin Orbit Coupling}

\author{Daniel Shaffer}

\affiliation
{
Department  of  Physics,  Emory  University,  400 Dowman Drive, Atlanta,  GA  30322,  USA
}

\author{Dmitry V. Chichinadze}
\affiliation
{
School of Physics and Astronomy, University of Minnesota, Minneapolis, MN 55455, USA
}

\begin{abstract}
A lot of attention has been drawn to superconductivity in UTe\(_2\), with suggestions of time-reversal symmetry breaking triplet chiral superconducting order parameter. The chirality of the order parameter has been attributed to an accidental near degeneracy of two superconducting components belonging to 1D irreps \(B_{2u}\) and \(B_{3u}\) of the relevant \(D_{2h}\) point group, and it has been argued that the chiral \(B_{2u}+iB_{3u}\) combination is selected by ferromagnetic fluctuations. In this work we present a possible explanation of the near-degeneracy as a result of an accidental \(C_4\) symmetry of the band structure, with the superconducting order parameter belonging the 2D \(E_u\) irrep of \(D_{4h}\) that uniquely descends to the sought after \(B_{2u}+iB_{3u}\) combination. We show that the \(C_4\) symmetry is emergent at the level of the interactions using a renormalization group calculation and argue that the chiral combination of the order parameter is favored when spin-orbit coupling is added to the model.
\end{abstract}

\maketitle

\section{Introduction}

Recent experiments suggest that UTe\(_2\) may be a chiral triplet superconductor (SC) \cite{RanTripletDiscovery19,Vidya20,PaglioneSpecificHeatNodes19} (see also \cite{AokiYanase21} for a recent review). In particular, SC has been observed in extremely high magnetic fields for all field directions, including re-entrant SC \cite{RanReentrant19} possibly due to the orbital effect \cite{Lebed20, Mineev20, ParkYBKim20}, and there are strong indications of spontaneous time reversal symmetry (TRS) breaking \cite{RanTripletDiscovery19,Vidya20,PaglioneSpecificHeatNodes19,WeiAgterbergSchmalian22}. However, UTe\(_2\) is strongly orthorombic with a \(D_{2h}\) point group symmetry which has only one-dimensional irreducible representations (irreps). This poses a problem since the realization of chiral superconductivity requires a gap function with multiple degenerate components (meaning each component has the same critical temperature in the linearized gap equation), which is only guaranteed to happen for higher dimensional irreps \cite{Scheurer17}. To circumvent this issue, it has been proposed that there is an accidental near degeneracy of two gap components belonging to two different 1D irreps of \(D_{2h}\) that effectively act as a single 2D irrep.

In particular, the \(B_{2u}\) and \(B_{3u}\) irreps have been proposed as their symmetry closely matches experimental observations \cite{AgterbergDFT21, AgterbergHeatCapacity2Tranistions20, XuDFT19, WeiAgterbergSchmalian22}. Magnetic fluctuations have also been invoked to explain why the chiral \(B_{2u}+iB_{3u}\) combination, corresponding to $p \pm ip$ pairing in terms of angular harmonics, is preferred over the time-reversal symmetry preserving \(B_{2u}+B_{3u}\) combination \cite{AgterbergHeatCapacity2Tranistions20,WeiAgterberg21,WalkerSamokhin02}. The mechanism is consistent with first-principles calculations that indicate that the Uranium $f$-orbital is localized, giving rise to ferromagnetic fluctuations \cite{IshizukaYanase21,XuDFT19}.
The choice of two order parameters with different symmetries is also supported by recent observation of two jumps in heat capacity measurements indicating two phase transitions \cite{AgterbergHeatCapacity2Tranistions20}. This is consistent with the fact that the accidental degeneracy is not expected to be exact. However, no apparent underlying reason for why those two particular irreps are present and are nearly degenerate has been offered.

In this work we propose a simple possible source of the near degeneracy and illustrate it using a simple minimal band structure model. The model is not intended to closely match the band structure obtained in experiments and numerical simulations. However, it captures what we argue is the main qualitative feature of UTe\(_2\): the quasi-1D nature of the Fermi surfaces.
This is motivated by DFT calculations \cite{IshizukaYanase21,XuDFT19} and ARPES data \cite{WrayARPES20} suggesting that the itinerant electrons are largely constrained to perpendicular 1D U and Te chains. Our model therefore consists of two arrays of perpendicular wires made of U and Te atoms correspondingly. It is mathematically similar to models that have been used in the context of proximitized twisted quantum wires \cite{Franz2021}, crossed sliding Luttinger liquids \cite{MukhopadhyayLubensky01_1,MukhopadhyayLubensky01_2}, two-legged ladder models \cite{SheltonPRB1996, GiamarchiPRB1999, LinBalentsFisher97, LinBalentsFisher98}, and the quasi-1D model for Sr\(_2\)RuO\(_4\) \cite{raghuSRO13}.

In the absence of interactions, we first show that the model has an \emph{accidental} approximate \(D_{4h}\) symmetry of the Bogolyubov-de Gennes Hamiltonian which \emph{does} have a 2D irrep \(E_u\). Under the breaking of the accidental \(C_4\) symmetry \(D_{4h}\rightarrow D_{2h}\) and the two components of the \(E_u\) irrep descend to \(B_{2u}\) and \(B_{3u}\) irreps, in agreement with observations. Adding the interactions that include spin fluctuations, we find that under the renormalization group (RG) flow the intra-chain coupling constants flow to the same values at least for some range of the bare coupling constant, even if the bare constants break the accidental \(C_4\) symmetry. The result is an \emph{emergent} \(D_{4h}\) symmetry. A related phenomenon has been conjectured for the two-channel Kondo lattice \cite{Coop_Coleman_1998,Tsvelick1984,Andrei_Kondo,Coleman_Two_Channel}, though it was not supported by RG calculations \cite{FabrizioNozieres95}. The RG flow for our model also leads to a triplet instability belonging to the desired \(E_u\) irrep. Finally, we show that the free energy is minimized by a chiral \(p+ip\) paired state when spin-orbit coupling is included. Our approach thus provides a microscopic explanation of the observed time reversal symmetry breaking triplet pairing and two superconducting transitions associated with chiral structure of the order parameter.

The structure of the paper is as follows. In Sec. \ref{sec:mod} we introduce the free-fermion model and discuss the structure of Fermi surfaces. In Sec. \ref{sec:BdG} we discuss the approximate symmetry of BdG Hamiltonian and its effect on SC order parameters. Sec. \ref{sec:int} is dedicated to the RG analysis of interactions and the emergence of effective $C_4$ symmetry. We show that the ground state is a chiral superconductor using Landau functional for SC order parameters in Sec. \ref{sec:chiral}. We discuss our results and conclude in Sec. \ref{sec:disc}.

\section{Minimal Model and Fermi Surfaces}
\label{sec:mod}

\begin{figure}[h]
\centering
\includegraphics[width=0.99\linewidth]{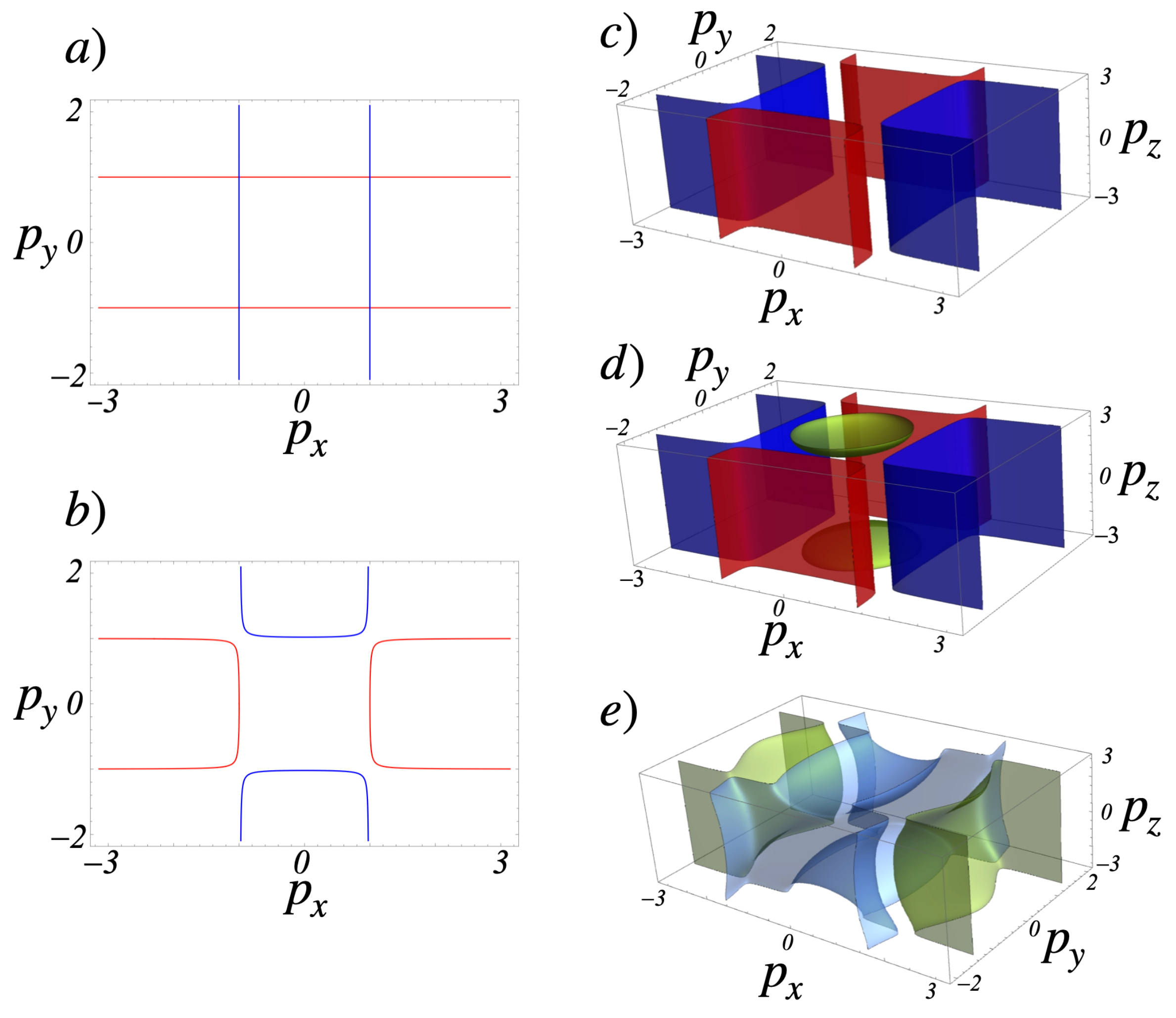}
\caption{Fermi surfaces of UTe\(_2\) in various approximations. a) Model without SOC assuming pure 1D dispersion along U (blue curves) and Te (red curves). We also show the ordering vectors corresponding to peaks in intra-atomic U (red arrows), intra-atomic Te (blue arrows), and inter-atomic (purple arrows) susceptibilities. b) and c) Quasi-1D model with SOC shown in 2D and 3D Brillouin zones. d) Quasi-1D model with an additional \(Z\) electron pocket without SOC between \(d\) and \(f\) U electrons. e) Same as d) but including SOC between \(d\) and \(f\) U electrons. Note that the \(Z\) electron pocket hybridizes with the \(X\) electron pocket as a result, forming a doughnut shape (actually a double doughnut once the walls of the BZ are identified). Cf. with DFT calculations and ARPES data \cite{XuDFT19,WrayARPES20,AgterbergDFT21}.
 }
\label{FS}
\end{figure}

The actual Fermi surfaces of UTe\(_2\) have been partially measured in ARPES experiments \cite{WrayARPES20}, but there are still a few issues that have not been resolved, in particular whether or not a pocket is present at the \(Z\) point in Brillouin zone (BZ). Nevertheless, the known features of the Fermi surfaces can be understood in a sequence of approximations (see Fig. \ref{FS}). The relevant degrees of freedom originate from the \(p\) orbital holes of Te atoms and one electron each from the \(d\) and \(f\) orbitals of the U atoms. The U atoms form double chains while Te atoms form single chains running in the orthogonal direction. The role of the \(f\) electrons is the main unresolved question in the literature: in DFT calculations with intermediate interactions, they form the \(Z\) pockets, but at larger \(U\) the pockets are gapped out, possibly due to the Kondo physics \cite{XuDFT19,IshizukaYanase21}. We will mostly ignore the \(f\) electrons for simplicity, though they likely play an important indirect role in producing ferromagnetic spin fluctuations that we do include later in our interaction model below.

Neglecting the \(f\) electrons, the U double chains can be thought of as a single chain, and since the separation between the chains is much larger than the separation between atoms within the chain, to zeroth order the dispersions are 1D and can be described by a \(2\times2\) \(\mathbf{k\cdot p}\) Hamiltonian (not including spin):
\[\mathcal{H}_0=\left(\begin{array}{cc}
    \frac{p^2_x}{2m_U}-\mu_U &0  \\
    0 & -\frac{p^2_y}{2m_{Te}}-\mu_{Te}
\end{array}\right)\]
In the simplest case which we will adopt \(m_U=m_{Te}\) and \(\mu_U=-\mu_{Te}=\mu\). This is of course somewhat far away from the real system, but matches the qualitative features of the ARPES Fermi surface data surprisingly well. The Fermi surfaces in this approximation are simply straight lines, orthogonal for the electrons and holes (see Fig. \ref{FS} a)).

\subsection*{Corrections to the minimal model from spin-orbit coupling and the $Z$-pocket}

There are several properties of UTe\(_2\) that the minimal model does not capture that we discuss here for completeness.
First of all, an \(\mathbf{L\cdot S}\) type spin-orbit coupling (SOC) is present in the real system. The SOC hybridizes the \(p\) and \(d\) orbitals, splitting the Fermi surfaces into an electron and a hole pocket centered at the \(Y\) and \(X\) points. For moderate values of SOC, the splitting is small and the pockets are nearly rectangular, as shown in Fig. \ref{FS} b) and c). There are several symmetry allowed SOC terms, but to capture the qualitative effect it is enough to include one:
\[\mathcal{H}_1=\mathcal{H}_0+\alpha\sigma^0\tau^x=\left(\begin{array}{cc}
    \frac{p^2_x}{2m}-\mu & \alpha  \\
    \alpha & -\frac{p^2_y}{2m}+\mu
\end{array}\right)\label{HwSOC}\]
where \(\sigma\) is the spin Pauli matrix, \(\tau\) is the Pauli matrix in the space of U/Te degrees of freedom, and \(\alpha\) is the SOC strength. Note we can take the SOC term to be nominally spin-independent due to inversion symmetry.

As mentioned above, we also neglect the \(f\) electron pocket at \(Z\) point in BZ in our model, but in principle it can be included, as shown in Fig. \ref{FS} d). If present, it is in general hybridized with the \(Y\) pocket due to SOC between the \(d\) and \(f\) uranium orbitals, resulting in a doughnut shaped Fermi surface (see Fig. \ref{FS} e)). We assume that even if the \(Z\) pocket is present, it participates only weakly in the superconducting condensate, and so can be neglected. In contrast, though we can initially neglect the SOC between \(p\) and \(d\) orbitals, we will show in Sec. \ref{sec:chiral} that it has an important role in determining the relative phase between the two components of the superconducting order parameter.

\section{Accidental Approximate Symmetry of the Bogolyubov-de Gennes Hamiltonian}
\label{sec:BdG}

We include superconductivity by considering the Bogolyubov-de Gennes (BdG) Hamiltonian in the absence of spin-orbit coupling
\begin{align}
\mathcal{H}_{Bdg}&=\left(\begin{array}{cc}
    \mathcal{H}_0(\mathbf{p}) & \hat{\Delta}(\mathbf{p})  \\
    \hat{\Delta}^\dagger(\mathbf{p})  & -\mathcal{H}^*_0(-\mathbf{p})
\end{array}\right)\nonumber\\
&=\left(\begin{array}{cccc}
    \frac{p^2_x}{2m}-\mu &0 & \Delta_U & 0 \\
    0 & -\frac{p^2_y}{2m}+\mu & 0 & \Delta_{Te} \\
   \Delta_U^\dagger & 0 & -\frac{p^2_x}{2m}+\mu &0  \\
   0 & \Delta_{Te}^\dagger & 0 & \frac{p^2_y}{2m}-\mu  \\
\end{array}\right)
\end{align}
where \(\Delta_U\) and \(\Delta_{Te}\) are \(2\times2\) matrices in spin space. Note that due to inversion symmetry, the pairing is predominantly between two uranium electrons or two tellurium holes (i.e. between opposite sides of the Fermi surfaces), and the pairing between U and Te can be neglected as a result. In the BdG formalism we introduce the Nambu spinors \(\Psi_{\tau\sigma}(\mathbf{p})=\left(\psi_{\tau\sigma}(\mathbf{p}),\psi^\dagger_{\tau\sigma}(-\mathbf{p})\right)\) (\(\tau\) labels U or Te), resulting in a two-fold redundancy. This is accounted for by the anti-unitary particle-hole symmetry (PHS) which acts as \(\mathcal{C}=\varsigma^x\mathcal{K}\) where \(\varsigma^x\) is a Pauli matrix acting on the new particle/hole degrees of freedom and \(\mathcal{K}\) is complex conjugation. PHS in particular requires \(\hat{\Delta}(\mathbf{p})=-\hat{\Delta}^\dagger(-\mathbf{p})\).

We make the following observation: in the normal state with \(\hat{\Delta}=0\), the BdG Hamiltonian has a new unitary symmetry that is not present in the original Hamiltonian. The new symmetry is \(C_4\tau^x\varsigma^x\), where \(C_4\) is a four-fold rotation symmetry taking \(p_x\rightarrow p_y\) and \(p_y\rightarrow -p_x\). This is intuitively clear: in the BdG Hamiltonian we introduce additional redundant copies of the Fermi surfaces but of opposite characters (electron instead of hole and vice versa). The full symmetry of the BdG Hamiltonian in the simplest version of our model is therefore not \(D_{2h}\times \mathcal{P}\) (\(\mathcal{P}\) being the PHS symmetry group), but rather \(D_{4h}\times \mathcal{P}\). Of course the symmetry is only exact in our over-simplified model, but it remains an approximate symmetry as long as the neglected terms are not too large (we address some possible sources of such terms in Appendix \ref{AppendixA}).

Since the effective point group is \(D_{4h}\), the gap functions have to be classified according to irreducible representations (irreps) of \(D_{4h}\), not \(D_{2h}\). We note that \(D_{4h}\) has two 2D irreps, \(E_g\) and \(E_u\), corresponding to singlet and triplet pairing respectively. Since the symmetry is only approximate, the irreps of \(D_{4h}\) descend into irreps of \(D_{2h}\). For \(E_u\), the two components \(E_{u}^{(1)}\) and \(E_{u}^{(2)}\) descend into two different 1D irreps of \(D_{2h}\), \(E_{u}^{(1)}\rightarrow B_{2u}\) and \(E_{u}^{(2)}\rightarrow B_{3u}\) \cite{Dresselhaus08}. Therefore the chiral \(E_u\) phase, if established, descends uniquely into the \(B_{2u}+i B_{3u}\) phase, explaining the experimental observations and ruling out other proposed combinations (e.g. the non-unitary pairing proposed in \cite{Nevidomskyy20}). Moreover, since the \(C_4\) symmetry is broken, we expect the degeneracy between \(B_{2u}\) and \(B_{3u}\) also to be inexact, resulting in one of those two channels having higher \(T_c\), in agreement with the two jumps seen in the specific heat data \cite{AgterbergHeatCapacity2Tranistions20}.

A valid objection to the argument above is that it only holds for the non-interacting part of the Hamiltonian. The interactions, on the other hand, only have to respect the \(D_{2h}\) symmetry and not necessarily the accidental \(D_{4h}\). As we show in the following section, however, the accidental \(C_4\) symmetry 
of the non-interacting Hamiltonian
leads to an \emph{emergent} \(C_4\) symmetry on the level of the interaction Hamiltonian within the RG approach.

\section{Emergent \(C_4\) Symmetry of the Interactions Under RG Flow}
\label{sec:int}

In this section we address the issue of the absence of \(C_4\) symmetry at the interaction level by showing that the \(C_4\) symmetry can be emergent within the RG paradigm. The key to the result is the quasi-1D nature of the system.  Assuming that the interactions are dominant between sites within the chains and negligible between neighboring chains as well as between perpendicular chains, the interactions within the chains are essentially one-dimensional, and we can write down two sets of decoupled RG equations for each set of parallel chains. We label those chains \(H\) and \(V\) for horizontal (U) and vertical (Te). Note that in 1D one can solve the interaction problem exactly by bosonization, which has also been done for crossed wire networks to obtain the so-called crossed sliding Luttinger liquid \cite{MukhopadhyayLubensky01_1,MukhopadhyayLubensky01_2}. Before introducing the coupling between the \(H\) and \(V\) wires, we assume that they are in the Luttinger liquid regime, i.e. all the RG flows are at most marginal (or else irrelevant). The bare coupling constants of \(H\) and \(V\) wires are otherwise not assumed to be related, and so explicitly break the \(C_4\) symmetry. We then include inter-chain coupling between \(H\) and \(V\) wires as a perturbation and find that it results in an instability (i.e. a relevant flow in RG). The general argument is that since the instability is driven by the infinitesimal inter-chain interactions, the final fixed trajectory of the RG flow does not sensitively depend on the particular choice of the intra-chain coupling constants (at least for some range), and is therefore \(C_4\) symmetric. We verify numerically that the \(H\) and \(V\) coupling constants are equal for some choices of the bare interactions, i.e. there is an emergent \(C_4\) symmetry.

\begin{figure}[h]
\includegraphics[width=0.8\linewidth]{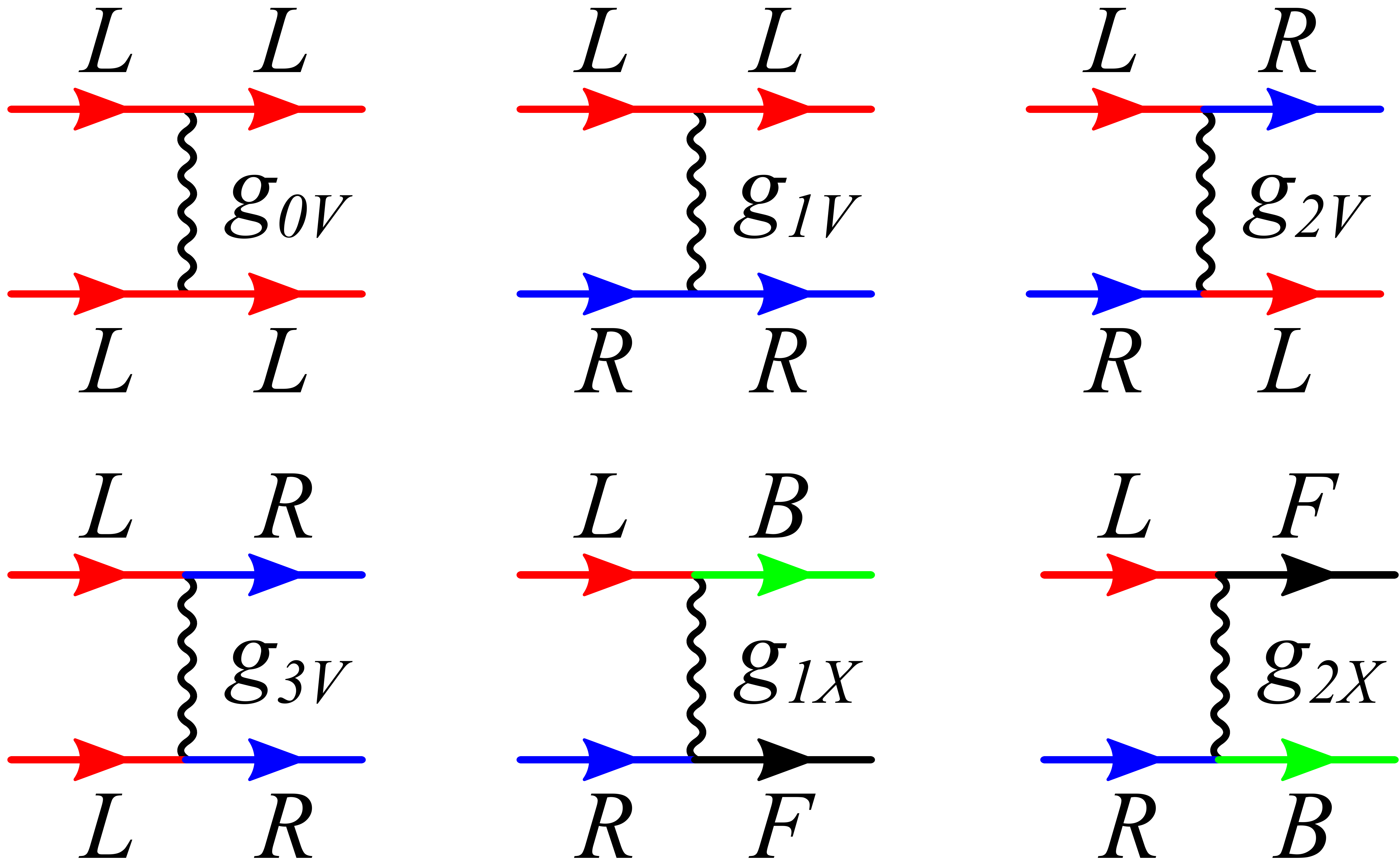}
\centering{}\caption{Coupling constants \(g_{nA}\) with \(g=u,J^{(j)}\), \(n=0,1,2,3\), \(A=V,X\) (the diagrams for \(A=H\) are similar with \(L\rightarrow B\) and \(R\rightarrow F\)).
} 
\label{Fig:couplings}
\end{figure}

In 1D, the Fermi surfaces are two points, and as usual we label them \(L\) and \(R\) (left and right) for the horizontal chains and \(B\) and \(F\) (bottom and front) for the vertical chains. The interactions within the vertical chains then have the following form (see Fig. \ref{Fig:couplings}):
\begin{align}
    H_{V}&=\frac{1}{2}\sum u_{0V}c^\dagger_{A\alpha}\delta_{\alpha\alpha'}c_{A\alpha'}c^\dagger_{A\beta}\delta_{\beta\beta'}c_{A\beta'}+\nonumber\\
    &+\frac{1}{2}\sum u_{1V}c^\dagger_{B\alpha}\delta_{\alpha\alpha'}c_{B\alpha'}c^\dagger_{F\beta}\delta_{\beta\beta'}c_{F\beta'}+\nonumber\\
    &+\frac{1}{2}\sum u_{2V}c^\dagger_{B\alpha}\delta_{\alpha\alpha'}c_{F\alpha'}c^\dagger_{F\beta}\delta_{\beta\beta'}c_{B\beta'}+\nonumber\\
    &+\frac{1}{2}\sum u_{3V}c^\dagger_{B\alpha}\delta_{\alpha\alpha'}c_{F\alpha'}c^\dagger_{B\beta}\delta_{\beta\beta'}c_{F\beta'}+\nonumber\\
    &+\frac{1}{2}\sum J^{(j)}_{0V}c^\dagger_{A\alpha}\sigma^j_{\alpha\alpha'}c_{A\alpha'}c^\dagger_{A\beta}\sigma^j_{\beta\beta'}c_{A\beta'}+\nonumber\\
    &+\frac{1}{2}\sum J^{(j)}_{1V}c^\dagger_{B\alpha}\sigma^j_{\alpha\alpha'}c_{B\alpha'}c^\dagger_{F\beta}\sigma^j_{\beta\beta'}c_{F\beta'}+\nonumber\\
    &+\frac{1}{2}\sum J^{(j)}_{2V}c^\dagger_{B\alpha}\sigma^j_{\alpha\alpha'}c_{F\alpha'}c^\dagger_{F\beta}\sigma^j_{\beta\beta'}c_{B\beta'}+\nonumber\\
    &+\frac{1}{2}\sum J^{(j)}_{3V}c^\dagger_{B\alpha}\sigma^j_{\alpha\alpha'}c_{F\alpha'}c^\dagger_{B\beta}\sigma^j_{\beta\beta'}c_{F\beta'}+h.c.
\end{align}
(\(A\) is summed over \(B\) and \(F\) while \(j\) is summed over \(x,y,z\)). Within the horizontal chains we have similarly (with \(A\) summed over \(L\) and \(R\) instead):
\begin{align}
    H_{H}&=\frac{1}{2}\sum u_{0H}d^\dagger_{A\alpha}\delta_{\alpha\alpha'}d_{A\alpha'}d^\dagger_{A\beta}\delta_{\beta\beta'}d_{A\beta'}+\nonumber\\
    &+\frac{1}{2}\sum u_{1H}d^\dagger_{L\alpha}\delta_{\alpha\alpha'}d_{L\alpha'}d^\dagger_{R\beta}\delta_{\beta\beta'}d_{R\beta'}+\nonumber\\
    &+\frac{1}{2}\sum u_{2H}d^\dagger_{L\alpha}\delta_{\alpha\alpha'}d_{R\alpha'}d^\dagger_{R\beta}\delta_{\beta\beta'}d_{L\beta'}+\nonumber\\
    &+\frac{1}{2}\sum u_{3H}d^\dagger_{L\alpha}\delta_{\alpha\alpha'}d_{R\alpha'}d^\dagger_{L\beta}\delta_{\beta\beta'}d_{R\beta'}+\nonumber\\
    &+\frac{1}{2}\sum J^{(j)}_{0H}d^\dagger_{A\alpha}\sigma^j_{\alpha\alpha'}d_{A\alpha'}d^\dagger_{A\beta}\sigma^j_{\beta\beta'}d_{A\beta'}+\nonumber\\
    &+\frac{1}{2}\sum J^{(j)}_{1H}d^\dagger_{L\alpha}\sigma^j_{\alpha\alpha'}d_{L\alpha'}d^\dagger_{R\beta}\sigma^j_{\beta\beta'}d_{R\beta'}+\nonumber\\
    &+\frac{1}{2}\sum J^{(j)}_{2H}d^\dagger_{L\alpha}\sigma^j_{\alpha\alpha'}d_{R\alpha'}d^\dagger_{R\beta}\sigma^j_{\beta\beta'}d_{L\beta'}+\nonumber\\
    &+\frac{1}{2}\sum J^{(j)}_{3H}d^\dagger_{L\alpha}\sigma^j_{\alpha\alpha'}d_{R\alpha'}d^\dagger_{L\beta}\sigma^j_{\beta\beta'}d_{R\beta'}+h.c.
\end{align}
Finally, the perturbing interactions between the two chains are
\begin{align}
    H_{X}&=\frac{1}{2}\sum u_{1X}d^\dagger_{L\alpha}\delta_{\alpha\alpha'}c_{B\alpha'}d^\dagger_{R\beta}\delta_{\beta\beta'}c_{F\beta'}+\nonumber\\
    &+\frac{1}{2}\sum u_{2X}d^\dagger_{L\alpha}\delta_{\alpha\alpha'}c_{F\alpha'}d^\dagger_{R\beta}\delta_{\beta\beta'}c_{B\beta'}+\nonumber\\
    &+\frac{1}{2}\sum J^{(j)}_{1X}d^\dagger_{L\alpha}\sigma^j_{\alpha\alpha'}c_{B\alpha'}d^\dagger_{R\beta}\sigma^j_{\beta\beta'}c_{F\beta'}+\nonumber\\
    &+\frac{1}{2}\sum J^{(j)}_{2X}d^\dagger_{L\alpha}\sigma^j_{\alpha\alpha'}c_{F\alpha'}d^\dagger_{R\beta}\sigma^j_{\beta\beta'}c_{B\beta'}+h.c.
\end{align}

These are the only momentum-conservation allowed interactions that are relevant for the RG flows. \(u\)'s correspond to density-density interactions, while \(J\)'s correspond to spin fluctuations; we use \(g\) as a generic label for coupling constants of either type. We include spin fluctuations as they have been suggested to mediate the triplet superconductivity \cite{AgterbergHeatCapacity2Tranistions20, WeiAgterberg21, Duan20, KnafoAoki21, UTe2Rafael20}. \(g_0\) are intra-pocket interactions that as we will see do not flow under the RG and don't affect other flows, and so henceforth can be ignored. \(g_1\) are inter-pocket interactions, \(g_2\) are exchange interactions, and \(g_3\) are umklapp processes only allowed at half-filling (we will later drop the \(g_3\) terms and assume we are not at half-filling). Note that \(g_{nH}\) and \(g_{nV}\) coupling constants have to be real by time-reversal symmetry (TRS) and/or by mirror/rotation symmetries. The inter-chain interactions \(g_{nX}\) may, on the other hand, be complex due to the absence of \(C_4\) symmetry, but note that due to the \(xz\) and \(yz\) mirror symmetries that exchange \(B/F\) and \(L/R\) labels respectively while keeping the other two fixed, we must have \(g_{1X}=g_{2X}\), and we will therefore label them simply as \(u_X\) and \(J^{(j)}_X\) below.

\begin{figure}
\includegraphics[width=0.8\linewidth]{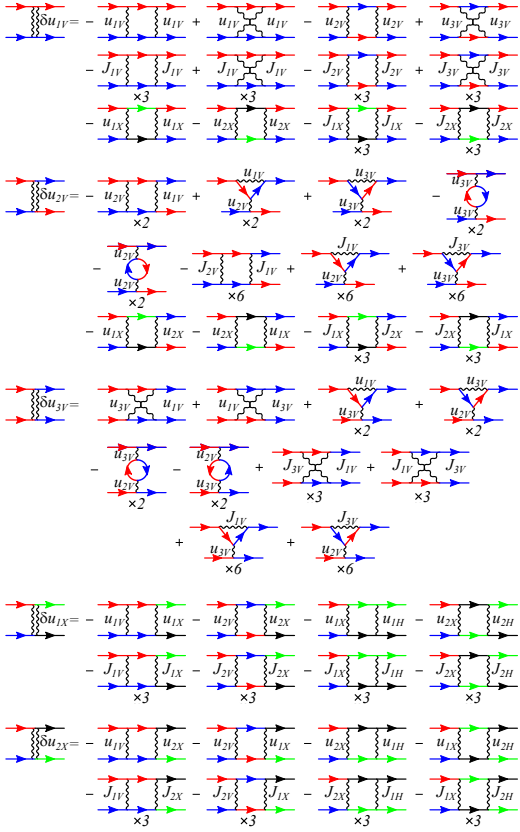}
\centering{}\caption{ Diagrammatic representation of RG equations for density-density interactions $u_i$. Note the corrections coming from the spin $J$ couplings that generate \(u_{nA}\) even if they are absent initially, see Appendix \ref{AppendixB}.  
} 
\label{Fig:RGflowU}
\end{figure}

\begin{figure}
\includegraphics[width=0.8\linewidth]{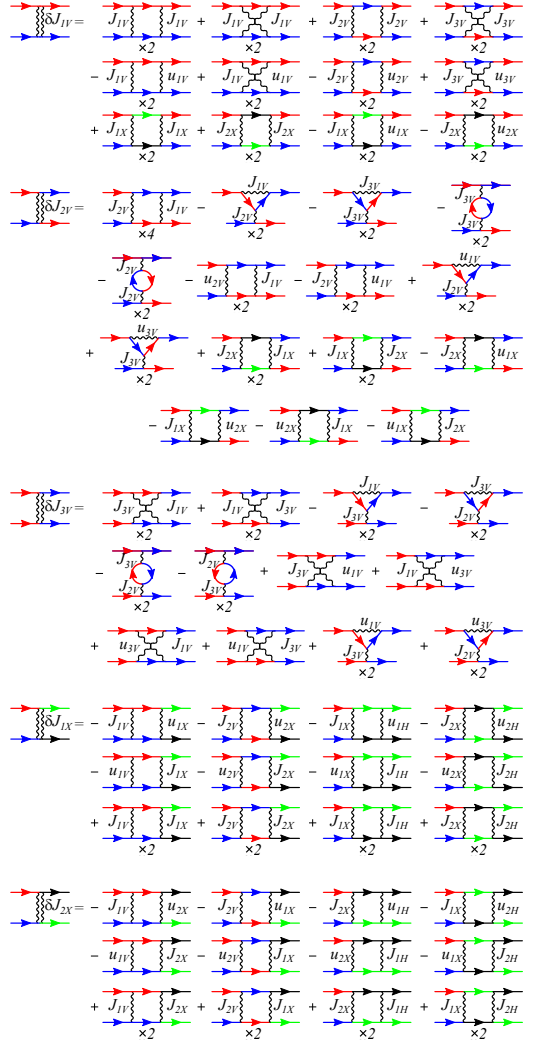}
\centering{}\caption{Diagrammatic representation of RG equations for spin-spin interactions $J_{nA}$. Note the corrections coming from the terms involving density-density $u_{nA}$.
} 
\label{Fig:RGflowJ}
\end{figure}

For simplicity, we will assume that the spin fluctuations are isotropic, \(J^{(x)}_{nA}=J^{(y)}_{nA}=J^{(z)}_{nA}=J_{nA}\) (\(n=0,1,2,3\), \(A=H,V\)). In that case we obtain the following RG equations (see Figs. \ref{Fig:RGflowU} and \ref{Fig:RGflowJ}):
\begin{align}\label{RGflowEq}
    \dot{u}_{1V}&=-u_{2V}^2+u_{3V}^2-3J_{2V}^2+3J_{3V}^2-\nonumber\\
    &-2\left|u_{X}\right|^2-6\left|J_{X}\right|^2\\
    \dot{u}_{2V}&=-2u_{2V}^2-6J_{1V}J_{2V}+6u_{3V}J_{3V}+6u_{2V}J_{1V}-\nonumber\\
    &-2\left|u_{X}\right|^2-6\left|J_{X}\right|^2\\
    \dot{u}_{3V}&=4u_{1V}u_{3V}-2u_{2V}u_{3V}+6u_{3V}J_{1V}+6u_{2V}J_{3V}+\nonumber\\
    &+6J_{1V}J_{3V}\\
    \dot{J}_{1V}&=2\left(-u_{2V}J_{2V}+u_{3V}J_{3V}+2J_{1V}^2+J_{2V}^2+J_{3V}^2\right)-\nonumber\\
    &-4\text{Re}\left[u_{X}^*J_{X}\right]+4\left|J_{X}\right|^2\\
    \dot{J}_{2V}&=2\left(u_{3V}J_{3V}-u_{2V}J_{1V}-J^{2}_{2V}+J_{1V}J_{2V}-2J_{3V}^2\right)+\nonumber\\
    &-4\text{Re}\left[u_{X}^*J_{X}\right]+4\left|J_{X}\right|^2\\
    \dot{J}_{3V}&=2\left(2u_{1V}J_{3V}+u_{3V}J_{1V}+u_{3V}J_{2V}\right)+\nonumber\\
    &+2\left(J_{1V}J_{3V}-3J_{2V}J_{3V}\right)\\
    \dot{u}_{X}&=-\left(u_{1V}+u_{2V}\right)u_{X}-u_{X}^*\left(u_{1H}+u_{2H}\right)-\nonumber\\
    &-3\left(J_{1V}+J_{2V}\right)J_{X}-3J_{X}^*\left(J_{1H}+J_{2H}\right)\\
    \dot{J}_{X}&=-\left(u_{1V}+u_{2V}\right)J_{X}-u_{X}\left(J_{1V}+J_{2V}\right)-\nonumber\\
    &-u_{X}^*\left(J_{1H}+J_{2H}\right)-\left(u_{1H}+u_{2H}\right)J_{X}^*+\nonumber\\
    &+2\left(J_{1V}+J_{2V}\right)J_{X}+2J_{X}^*\left(J_{1H}+J_{2H}\right)
\end{align}
with a similar set of equations for \(g_{nH}\). The details of the RG, including equations for non-isotropic spin-fluctuations, can be found in Appendix \ref{AppendixB}. Here we assumed for simplicity that the DOS of \(H\) and \(V\) chains are equal. If the DOS's are different we can recover the same equations by re-scaling the coupling constants as \(\tilde{g}_V=\nu_Vg_V\),\(\tilde{g}_H=\nu_Hg_H\) and \(\tilde{g}_X=\sqrt{\nu_V\nu_H}g_X\), so that the form of the equations is the same. For simplicity, we will keep assuming that the DOS's are equal, i.e. the non-interacting part of the Hamiltonian is \(C_4\) symmetric. We also henceforth set the umklapp processes \(g_{3A}\) to zero, as those generally lead to instabilities in 1D and we are interested in the Luttinger regime in the absence of inter-chain interactions \(g_X\). As mentioned above, this is justified on physical grounds as we do not expect both sets of chains to be exactly at half-filling.

\begin{figure*}
\includegraphics[width=0.95\linewidth]{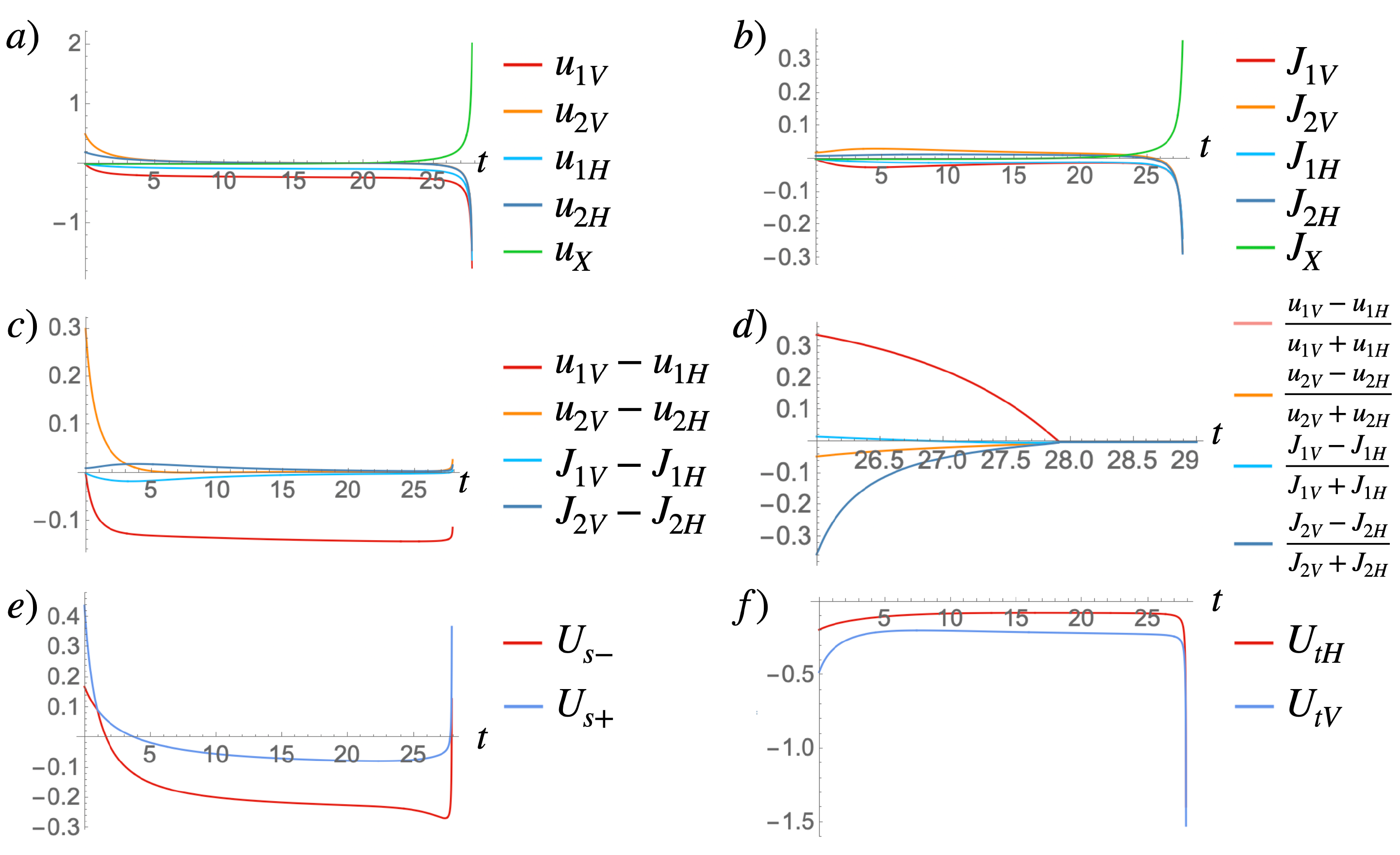}
\centering{}\caption{RG flow for the following values of the bare coupling constants: \(u_{2H}=0.2\), \(J_{2H}=0.01\), \(u_{2V}=0.5\), \(J_{2V}=0.02\), \(u_X=0.001\) and \(J_X=0.0003\). a) and b) show the flow of the density-density and spin-fluctuation mediated interactions respectively; the intra-chain coupling constants all flow to negative values while the inter-chain couplings \(u_X\) and \(J_X\) flow to positive values. c) shows the flow of the differences of intra-chain coupling between \(V\) and \(H\) chains, d) shows the same differences normalized by the sums. Note that the latter flow to zero at the instability, indicating an emergent \(C_4\) symmetry. e) and f) show the effective interactions \(U_{s\pm}\) in the singlet and \(U_{tV}\) and \(U_{tH}\) in the triplet pairing channels respectively (see Eqs. (\ref{Ut}-\ref{Us})), with negative/positive values corresponding to attraction/repulsion. We thus find that the triplet channel wins under the RG flow with the given bare coupling constants.
} 
\label{Fig:RGflowCurves}
\end{figure*}

The resulting RG flow is shown in Fig. \ref{Fig:RGflowCurves} (see caption and below for bare coupling constant values). As claimed, with the chosen values of the bare coupling constants the quantities \((g_{nV}-g_{nH})/(g_{nV}+g_{nH})\) flow to zero while \(g_{nV}+g_{nH}\) diverges, implying that \(g_{nV}=g_{nH}\) at the fixed point at infinity, i.e. there is an emergent \(C_4\) symmetry. This is true for a relatively wide range of the bare coupling constants, assuming no instabilities in the absence of inter-chain interactions \(g_X\) (there are also other fixed trajectories we observed along which \(g_{nV}=0\) or \(g_{nH}=0\)). 
This result matches the conjectured RG flow proposed for the related two-channel Kondo lattice \cite{Coop_Coleman_1998} -- the two initially different couplings become equal under the RG flow.

\subsection*{Vertex Flow Equations}

\begin{figure}[h]
\includegraphics[width=0.99\linewidth]{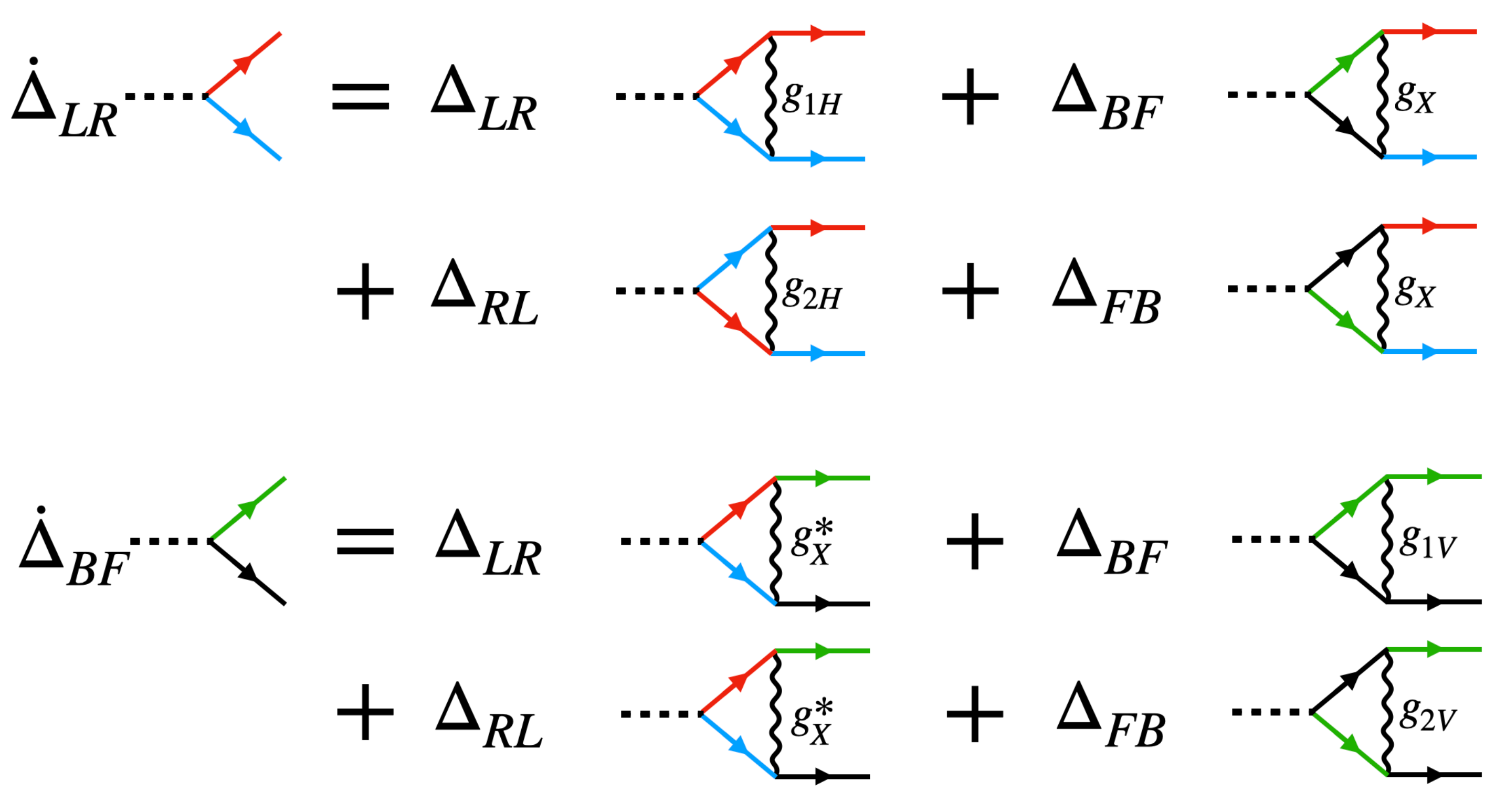}
\centering{}\caption{Diagrammatic representation of the SC vertex flow equations. The spin sum evaluation is presented in Appendix \ref{AppendixB}.
} 
\label{Fig:RGvertexFlow}
\end{figure}

Though we find an emergent \(C_4\) symmetry of the interactions, we also need to show that triplet superconductivity belonging to the \(E_u\) irrep is the leading instability. To show that, we introduce test vertices
\begin{align}
    \Delta_{LR}^{(\mu)}\left(\sigma^\mu i\sigma^y\right)_{\alpha\beta}d^\dagger_{\mathbf{p},L,\alpha}d^\dagger_{\mathbf{p},R,\beta}+h.c.\nonumber\\
    \Delta_{BF}^{(\mu)}\left(\sigma^\mu i\sigma^y\right)_{\alpha\beta}c^\dagger_{\mathbf{p},B,\alpha}c^\dagger_{\mathbf{p},F,\beta}+h.c\nonumber\\
    \Delta_{RL}^{(\mu)}\left(\sigma^\mu i\sigma^y\right)_{\alpha\beta}d^\dagger_{\mathbf{p},R,\alpha}d^\dagger_{\mathbf{p},L,\beta}+h.c.\nonumber\\
    \Delta_{FB}^{(\mu)}\left(\sigma^\mu i\sigma^y\right)_{\alpha\beta}c^\dagger_{\mathbf{p},F,\alpha}c^\dagger_{\mathbf{p},B,\beta}+h.c
\end{align}
and study their RG flow, including the competition between the singlet and triplet channels. The relevant diagrams are shown in Fig. \ref{Fig:RGvertexFlow}. Note that due to anti-commutation relations, we have the particle-hole symmetry relations
\begin{equation}
\begin{gathered}
    \Delta_{LR}^{(0)}=\Delta_{RL}^{(0)}=\Delta_H^{(0)}\\
    \Delta_{BF}^{(0)}=\Delta_{FB}^{(0)}=\Delta_V^{(0)}\\
    \Delta_{LR}^{(j)}=-\Delta_{RL}^{(j)}=\Delta_H^{(j)}\\
    \Delta_{BF}^{(j)}=-\Delta_{FB}^{(j)}=\Delta_V^{(0)}
    \label{gap_sym}
    \end{gathered}
\end{equation}
where \(j=x,y,z\) correspond to the triplet components and \(\mu=0\) is the singlet component of the gap functions. The RG flow equations of the vertices \(\Delta_H^{(\mu)}\) and \(\Delta_V^{(\mu)}\) are as follows \footnote{We again assume equal DOS's on \(H\) and \(V\) chains, but this difference can again be taken into account by re-scaling the vertices as \(\tilde{\Delta}_H^{(\mu)}=\sqrt{\nu_H}\Delta_H^{(\mu)}\) and \(\tilde{\Delta}_H^{(\mu)}=\sqrt{\nu_V}\Delta_V^{(\mu)}\). See Appendix \ref{AppendixA}}:
\begin{align}
    \dot{\Delta}_H^{(0)}&=-\left(u_{1H}+u_{2H}-3J_{1H}-3J_{2H}\right)\Delta_H^{(0)}-\nonumber\\
    &-2\left(u_{X}-3J_X\right)\Delta_V^{(0)}\nonumber\\
    \dot{\Delta}_V^{(0)}&=-\left(u_{1V}+u_{2V}-3J_{1V}-3J_{2V}\right)\Delta_V^{(0)}-\nonumber\\
    &-2\left(u^*_{X}-3J^*_X\right)\Delta_H^{(0)}\nonumber\\
    \dot{\Delta}_H^{(j)}&=-\left(u_{1H}-u_{2H}+J_{1H}-J_{2H}\right)\Delta_H^{(j)}\nonumber\\
    \dot{\Delta}_V^{(j)}&=-\left(u_{1V}-u_{2V}+J_{1V}-J_{2V}\right)\Delta_V^{(j)}
\end{align}
The relevant spin sums are shown Fig. \ref{Fig:vertex_spin_sum} in Appendix \ref{AppendixB}. Importantly,  the \(H\) and \(V\) triplet components are decoupled (thanks to the cancellation due to the mirror symmetries of the \(D_{2h}\) point group). As a result, if the \(C_4\) symmetry emerges at the RG fixed point, the triplet \(H\) and \(V\) channels become degenerate, meaning that they belong to a 2D irrep.

The effective pairing triplet interactions are simply
\[U_{tA}=u_{1A}-u_{2A}+J_{1A}-J_{2A}\label{Ut}\]
with \(A=H,V\). The effective singlet interactions are found by diagonalizing the matrix equation equation for the flow of \(\Delta_H^{(0)}\) and \(\Delta_V^{(0)}\), with
\[U_{s\pm}=\frac{1}{2}\left(U_{sVH}\pm\sqrt{U_{sVH}^2+16\left|u_{X}-3J_{X}\right|^2}\right)\label{Us}\]
where
\[U_{sVH}=u_{1H}+u_{1V}+u_{2H}+u_{2V}-3J_{1H}-3J_{1V}-3J_{2H}-3J_{2V}\]
Based on the flow equations, we then see that the triplet pairing is favored over singlet pairing either for large negative exchange \(u_{2A}\) interactions, or large negative (i.e. ferromagnetic) spin fluctuations \(J_{1A}\). However, the former is ruled out by our requirement that the interactions are marginal in the absence of inter-chain interactions \(g_X\), while the latter we find not to give rise to an emergent \(C_4\) symmetry. Instead, we find that the desired solution is obtained for a larger positive \(u_{2A}\) and a smaller but sizable positive (i.e. anti-ferromagnetic) \(J_{2A}\), with an even smaller \(u_X\) and an even smaller \(J_X\). 

As a concrete case we take the bare coupling constants to be \(u_{2H}=0.2\), \(J_{2H}=0.01\), \(u_{2V}=0.5\), \(J_{2V}=0.02\), \(u_X=0.001\) and \(J_X=0.0003\), and the rest zero. With these bare coupling constants the leading SC channel is indeed triplet, and since it is degenerate it belongs to the \(E_u\) irrep of \(D_{4h}\), as desired. Although the initial intra-chain spin-fluctuations are positive, i.e. anti-ferromagnetic, observe that under the RG flow they change sign and become ferromagnetic, which promotes the triplet instability. This may be consistent with the experimental observation of a nearby AFM instability in the presence of pressure \cite{UTe2Rafael20, Duan20, KnafoAoki21, IshizukaYanase21}.
We also note that although \(J_X\), which remains positive under the RG flow, is the smallest term, it is crucial for triplet SC since if it is set to zero we obtain a singlet instability.

\section{Origin of the chiral SC state and the role of SOC}
\label{sec:chiral}

Having established that the \(C_4\) symmetry emerges at the fixed trajectory of the RG flow along with a triplet superconducting order belonging to the 2D \(E_u\) irrep of \(D_{4h}\), there remains the question of the relative phase between the two gap functions \(\Delta_H\) and \(\Delta_V\) forming the two components of the irrep, since at the level of the one loop RG flow (or equivalently the linearized gap equation) any linear combination of the two is equivalent. The degeneracy is lifted by the fourth order term in the free energy \[\mathcal{F}^{(4)}=\beta_{\phi}\Delta_V^2(\Delta_H^{*})^2+c.c.=2\beta_\phi|\Delta_V|^2|\Delta_H|^2\cos2\phi\]
If \(\beta_\phi\) is positive, the free energy is minimized by \(\phi=\pm\pi/2\) leading to a time-reversal symmetry (TRS) breaking chiral order, whereas if \(\beta_\phi\) is negative the free energy is minimized by \(\phi=0,\pi\), leading to a TRS preserving phase (see, e.g., \cite{FernandesSchmalian19, MaitiChubukov13}; additional terms may instead favor a nematic combination but we assume that is not the case as this is not seen in experiment).

\begin{figure}[h]
\includegraphics[width=0.95\linewidth]{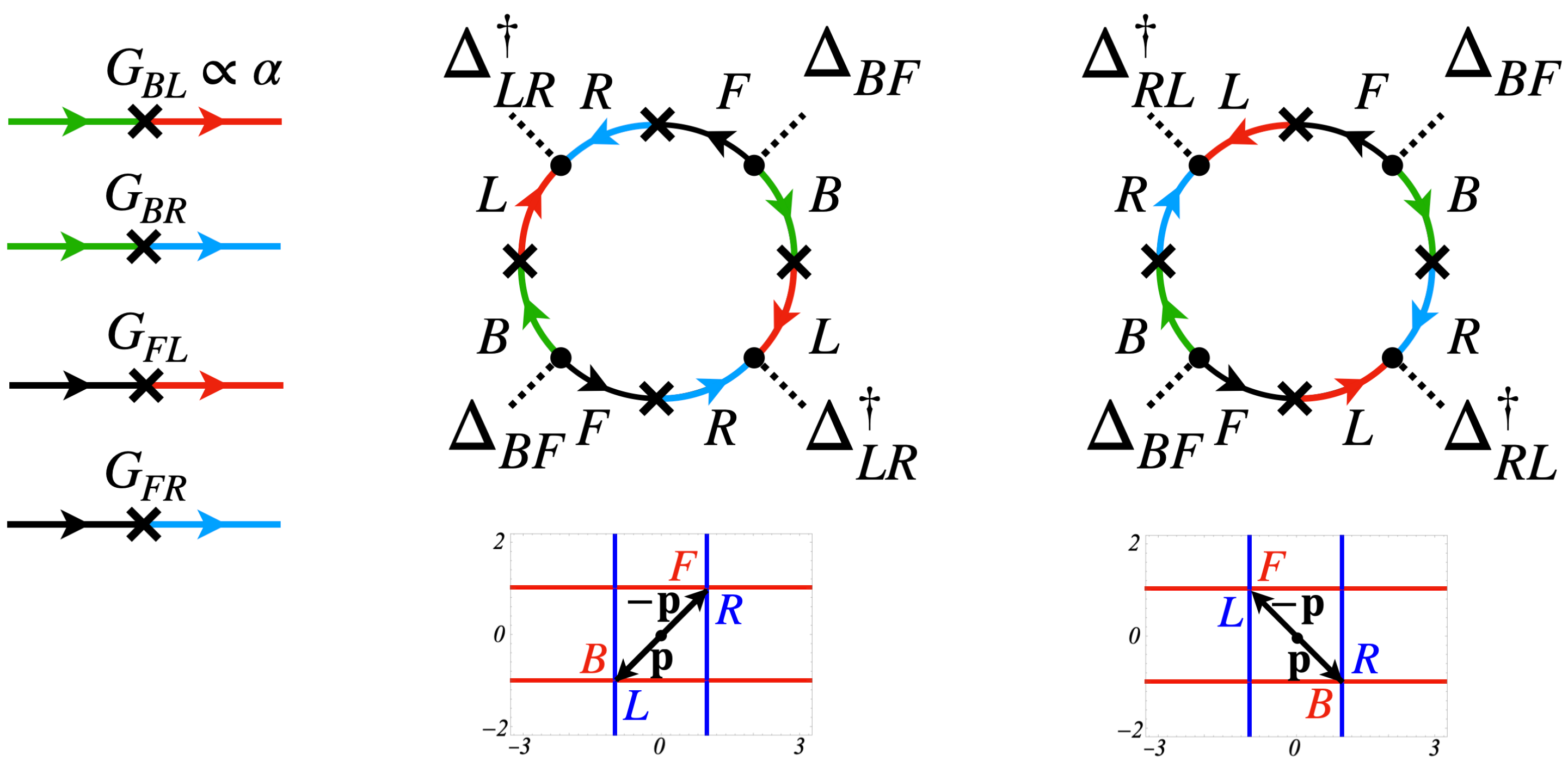}
\centering{}\caption{The fourth order diagrams that lift the degeneracy of the \(E_u\) irrep gap functions (two more diagrams are obtained by exchanging \(B\) and \(F\)). Note that the diagrams are only allowed when SOC is present, allowing \(V\)-chain electrons to change into \(H\)-chain electrons and vice-versa, giving rise to the off-diagonal Green's functions. For fermions close to the Fermi surface, the conversion can only happen around the four corners where the Fermi surfaces (without SOC) intersect. The kinematics are constrained such that if a \(B\) electron in the diagram is converted into an \(L\) electron, the electron with opposite momentum is converted from an \(F\) electron into an \(R\) electron, as indicated in the subplots below.
} 
\label{Fig:4th}
\end{figure}


The Feynman diagrams that contribute to this term must contain the two vertices \(\Delta_{BF}\) and two vertices \(\Delta_{LR}^\dagger\) (or vice-versa) connected by fermion propagators. This is not possible unless the propagator can change a fermion from a horizontal chain to one on the vertical chain. One possible way this can happen is through the inter-chain interactions, but one can check that for a triplet order parameter the resulting corrections vanish due to out of plane mirror symmetries. The only other possible way is through SOC: observe that the symmetry-allowed SOC in Eq. (\ref{HwSOC}) couples precisely Uranium and Tellurium electrons living on horizontal and vertical chains respectively. The corresponding Green's function is, to leading order in SOC,
\begin{align}
G(i\omega,\mathbf{p})&=\left(i\omega-\mathcal{H}_1(\mathbf{p})\right)^{-1}=\\
&=\left(\begin{array}{cc}
    \frac{1}{i\omega-\varepsilon_{U}(\mathbf{p})} & \frac{\alpha}{(i\omega-\varepsilon_{U}(\mathbf{p}))(i\omega-\varepsilon_{Te}(\mathbf{p}))}  \\
    \frac{\alpha}{(i\omega-\varepsilon_{U}(\mathbf{p}))(i\omega-\varepsilon_{Te}(\mathbf{p}))} &  \frac{1}{i\omega-\varepsilon_{Te}(\mathbf{p})}\nonumber
\end{array}\right)
\end{align}
Projecting the Green's function onto the \(A,A'=L,R,B,F\) patches, we find that the off-diagonal terms are thus given by \footnote{There are additional diagrams due to SOC and interactions that do not affect the analysis. For example, the lowest-order correction to free energy in $\alpha$ is $\sim \alpha^2\Delta^2$. The four possible contributions are:
$$F^{(2)} (\alpha) \sim \alpha^2 \left( \Delta_{LR} \Delta^*_{FB} + \Delta_{RL} \Delta^*_{FB} \right)+ c.c. $$
However, this sum vanishes for spin-triplet pairing because of the symmetry relations of gap functions \ref{gap_sym}.}
\[G_{AA'}(i\omega,\mathbf{p})=G_{A'A}(i\omega,\mathbf{p})=\frac{\alpha}{(i\omega-\varepsilon_{A}(\mathbf{p}))(i\omega-\varepsilon_{A'}(\mathbf{p}))}\]
The resulting fourth order diagram shown in Fig. \ref{Fig:4th} corresponds to a free energy term proportional to
\begin{align}
   \beta_{\phi}&=8T\sum_{n,\mathbf{p}}G^2_{BL}(i\omega,\mathbf{p})G^2_{FR}(-i\omega,-\mathbf{p})=\nonumber\\
   &=\sum_{n,\mathbf{p}}\frac{8T\alpha^4}{\left(\omega^2+\varepsilon_B^2(\mathbf{p})\right)^2\left(\omega^2+\varepsilon_L^2(\mathbf{p})\right)^2}\approx\nonumber\\
    &\approx\frac{\alpha^4\pi^2}{240T^5}\frac{1}{\left|v_{F,U}\right|\left|v_{F,Te}\right|}
\end{align}
where we took \(\varepsilon_B(\mathbf{p})=-\varepsilon_F(-\mathbf{p})\approx v_{F,Te} p_y\) and \(\varepsilon_R(\mathbf{p})=-\varepsilon_L(-\mathbf{p})\approx v_{F,U} p_x\); the momentum integral is done before the Matsubara sum. The factor of eight accounts for the spin summation and the fact that there are four contributing diagrams allowed by kinematics. Note that the SOC contributes significantly only around the four points where the quasi-1D Fermi surfaces intersect. The main point, however, is that \(\beta_\phi\) is positive, so that a chiral TRS-breaking order parameter is favored.

Note that the calculation of the fourth order term above does not depend at all on the form of the interactions, and so applies quite generally regardless of the specific pairing mechanism. In particular, our result is not incompatible with the previously proposed phenomenological explanation of the chirality of the order parameter via coupling to ferromagnetic fluctuations proposed in \cite{AgterbergHeatCapacity2Tranistions20, WeiAgterberg21}.

\section{Discussion}
\label{sec:disc}

In this work we have shown that the observed chiral triplet superconductivity in UTe\(_2\) can be explained by a combination of an accidental \(C_4\) symmetry (composed with a particle-hole symmetry) at the level of the non-interacting Hamiltonian, together with a resulting \emph{emergent} \(C_4\) symmetry of the interactions. Under the RG flow, a triplet SC order belonging to a 2D \(E_u\) irrep is established, and the chiral combination is selected when SOC is included. When the \(C_4\) symmetry is broken, the two components of the \(E_u\) irrep descend to a \(B_{2u}+i B_{3u}\) chiral combination of irreps of \(D_{2h}\), in agreement with experimental data \cite{AgterbergHeatCapacity2Tranistions20, WeiAgterberg21}.

The quasi-1D nature of the model plays a key role, allowing for a possibility that sans coupling between U and Te chains the system would be in a Luttinger liquid regime, with only marginal interactions in the RG flow. The inter-chain interactions then tilt the system towards the superconducting instability with an emergent \(C_4\) symmetry. One possible direction for a future study is to attempt a bosonized version of the calculation, since each 1D chain can then in principle be studied exactly  \cite{SheltonPRB1996, LinBalentsFisher97, LinBalentsFisher98, MukhopadhyayLubensky01_1, MukhopadhyayLubensky01_2, AiP1983,AndreiRMP,BosoBook}. Additionally, antiferromagnetism has been observed in samples under pressure \cite{UTe2Rafael20, Duan20, KnafoAoki21, IshizukaYanase21}, which may also be studied using the RG equations presented here and which may compete or be intertwined with the triplet superconducting state. In particular, the 1D chains may enter the Luther-Emery liquid phase in the limit of zero inter-chain interactions \cite{LutherEmery74}, resulting in an AFM or SDW instability. Umklapp processes that we ignored here may also play a significant role.

Within our model we generally neglected the \(f\)-electrons, which likely play a role in mediating the superconductivity via ferromagnetic fluctuations \(\mathbf{M}\sim f^\dagger_\alpha \sigma^z_{\alpha\beta}f_\beta\). Our model is not incompatible with this scenario as we do not postulate the origin of the bare couplings, though the RG equations do shed light on what type of microscopic interactions are compatible with the chiral triplet state. It may therefore be fruitful to study a more detailed microscopic model of the interactions within the RG framework presented here.

\section{Acknowledgements}

We thank P. Volkov, R. M. Fernandes, and L. Classen for useful discussions and A. Kamenev for referring us to literature. The authors are especially grateful to A. Chubukov for reading the manuscript and providing extremely useful feedback. DS was supported by startup funds at Emory University. DVC was supported by the Anatoly Larkin and Doctoral Dissertation Fellowships of the University of Minnesota. DVC also acknowledges the hospitality of KITP at Santa Barbara, where this project was initiated. The part of research done at KITP was supported in part by the National Science Foundation under Grant No. NSF PHY-1748958.

\appendix

\section{Sources of \(C_4\)-symmetry Breaking in the Free Hamiltonian}\label{AppendixA}

\begin{figure}
\includegraphics[width=0.99\linewidth]{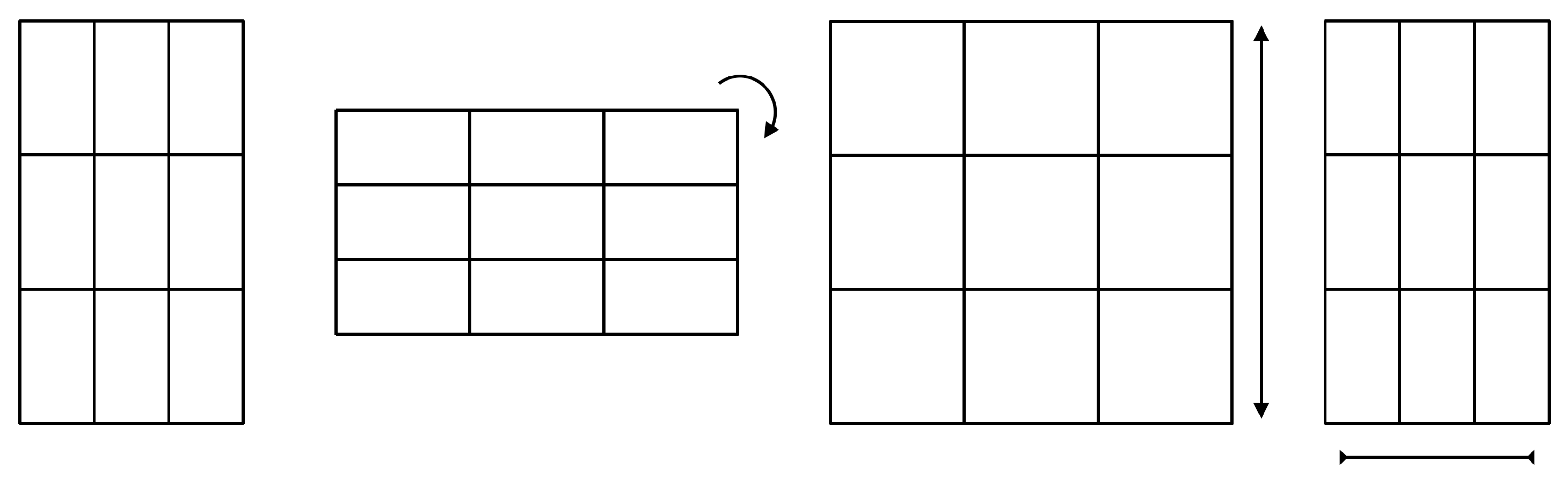}
\centering{}\caption{An extra symmetry of an orthorhombic system of wires: after a \(C_4\) rotation, rescaling along \(b\) and \(a\) brings the system of wires into the original system of wires. Alternatively, the image represents the Brillouin zone in momentum space and the Fermi surfaces. As long as the Fermi momenta of U and Te wires are in the same ratio as \(a\) and \(b\), the Fermi surfaces are mapped back to themselves under the same combination of symmetries.
} 
\label{Fig:C4rescaled}
\end{figure}

There are several possible sources of \(C_4\) symmetry breaking that we have neglected for simplicity, but two of those can be accommodated within our model by combining \(C_4\) with further symmetries. First of all, the \(a\) and \(b\) lattice parameters for UTe\(_2\) differ by a significant amount, around \(2/3\) \cite{WrayARPES20}. Though this does explicitly break \(C_4\) symmetry, for a system of crossed wires \(C_4\) combined with a re-scaling along \(a\) and \(b\) directions remains a symmetry, as shown in Fig. \ref{Fig:C4rescaled} (note that this is not the case for a general orthorhombic system and is a special property of the crossed-wires system; the rescaling can also be thought of as sliding the wires). As long as the Fermi momenta of the U and Te wires have the same ratio as \(a\) and \(b\), they also respect this symmetry (see Fig. \ref{Fig:C4rescaled}). The ratio of the Fermi momenta does appear to be close to \(a/b\) in ARPES data in \cite{WrayARPES20}, and as long as it is not too large there remains an effective \(C_4\)-like symmetry.


The second source of \(C_4\)-symmetry breaking that can also be accommodated in our model is the difference of densities of states of the U and Te chains. As mentioned in the main text, the difference can be absorbed into the definitions of the coupling constants and vertices, which can be rescaled as \(\tilde{g}_V=\nu_Vg_V\),\(\tilde{g}_H=\nu_Hg_H\), \(\tilde{g}_X=\sqrt{\nu_V\nu_H}g_X\), \(\tilde{\Delta}_H^{(\mu)}=\sqrt{\nu_H}\Delta_H^{(\mu)}\) and \(\tilde{\Delta}_H^{(\mu)}=\sqrt{\nu_V}\Delta_V^{(\mu)}\). We note incidentally that the difference in the domains of \(p_x\) and \(p_y\) discussed above similarly appears in the RG equations as numerical factors that can be absorbed into the density of states. In either case, there is therefore still an emergent \(C_4\)-like symmetry in the RG equations, but in addition to the \(C_4\) symmetry it includes a rescaling of the  \(d\) and \(c\) operators by \(d\rightarrow \sqrt{\nu_U/\nu_{Te}}d\) and \(c\rightarrow \sqrt{\nu_{Te}/\nu_{U}} c\). Importantly, this is an exact symmetry of the RG equations and our results remain valid modulo the rescaling.

\section{Details of the RG Calculation}\label{AppendixB}

\begin{figure}[t]
\includegraphics[width=0.99\linewidth]{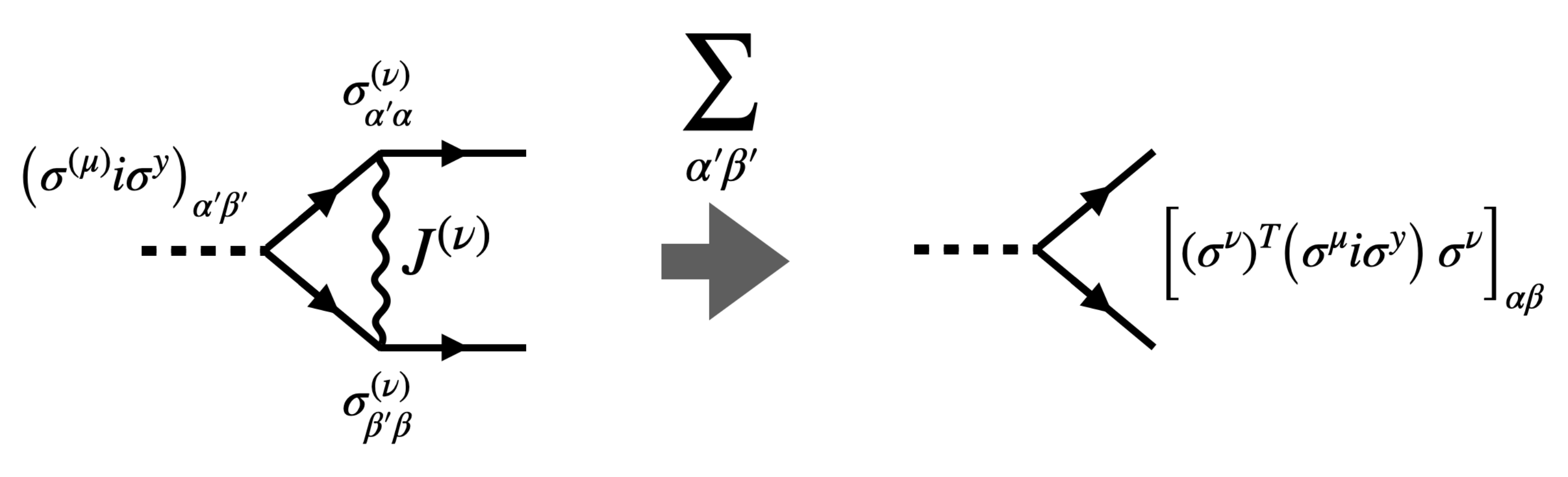}
\centering{}\caption{Spin sums involved in the one loop diagrams for the RG flow of the  particle-particle vertices. The diagram on the left contributes to the flow of the diagram on the right. \(\alpha'\) and \(\beta'\) are internal spin indices to be summed over
} 
\label{Fig:vertex_spin_sum}
\end{figure}

\begin{figure*}
\includegraphics[width=0.99\linewidth]{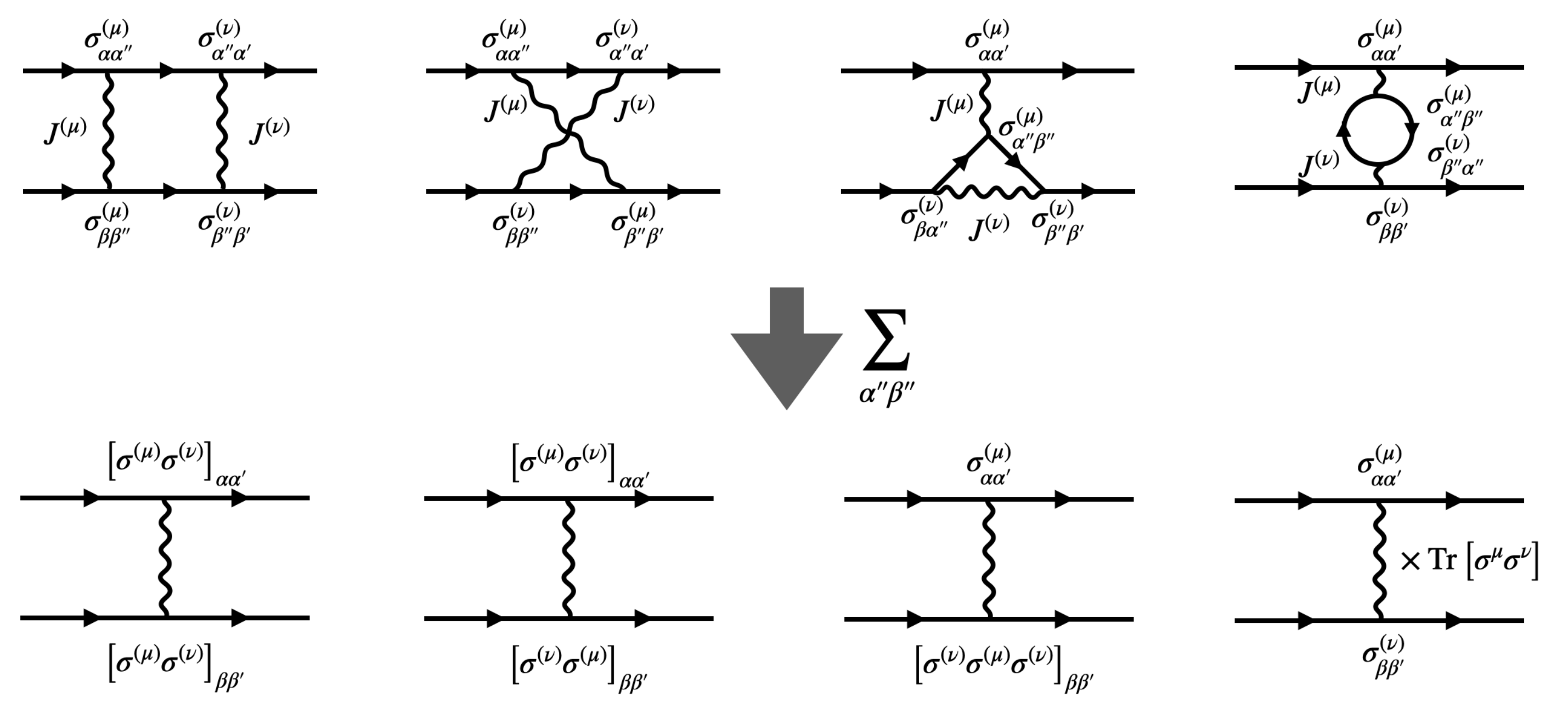}
\centering{}\caption{Spin sums involved in the one-loop diagrams for the RG flow of the coupling constants. Diagrams above contribute to the flow of the diagrams directly below. \(\alpha'\) and \(\beta'\) are internal spin indices to be summed over.
} 
\label{Fig:spin_sums}
\end{figure*}

In this appendix we show the details of the RG calculation. The one loop diagrams relevant for the RG flows are shown in Fig.'s \ref{Fig:RGflowU} and \ref{Fig:RGflowJ}. We need to consider the relevant spin summations, illustrated in Fig. \ref{Fig:spin_sums}. Here we use \(\sigma^0=\delta\) and \(J^{(0)}=u\), with \(\mu,\nu=0,x,y,z\). For the ladder diagrams we get
\[J^{(\mu)}J^{(\nu)}\left[\sigma^{(\mu)}\sigma^{(\nu)}\right]_{\alpha\alpha'}\left[\sigma^{(\mu)}\sigma^{(\nu)}\right]_{\beta\beta'}\]
For the crossed ladder diagram we get
\[J^{(\mu)}J^{(\nu)}\left[\sigma^{(\mu)}\sigma^{(\nu)}\right]_{\alpha\alpha'}\left[\sigma^{(\nu)}\sigma^{(\mu)}\right]_{\beta\beta'}\]
For the bubble diagram we get
\[J^{(\mu)}J^{(\nu)}\text{Tr}\left[\sigma^{(\mu)}\sigma^{(\nu)}\right]\sigma^{(\mu)}_{\alpha\alpha'}\sigma^{(\nu)}_{\beta\beta'}\]
(this accounts for the usual factor of two, and note that \(\text{Tr}\left[\sigma^{(\mu)}\sigma^{(\nu)}\right]=2\delta_{\mu\nu}\)). Finally, for the ``penguin" diagram we have
\[J^{(\mu)}J^{(\nu)}\left[\sigma^{(\mu)}\sigma^{(\nu)}\sigma^{(\mu)}\right]_{\alpha\alpha'}\sigma^{(\nu)}_{\beta\beta'}\]
with the \(\nu\) vertex being on the bottom (similarly for the upside down ``penguin" diagram). For completeness, Fig. \ref{Fig:vertex_spin_sum} shows the spin sums for the one loop vertex correction.


\begin{widetext}

Using these, we find the following RG flow equations for the coupling constants within the \(H\) chain (see Fig.'s \ref{Fig:RGflowU} and \ref{Fig:RGflowJ}):
\begin{align}
    \dot{u}_{1H}&=-u_{2H}^2+u_{3H}^2-\left|\mathbf{J}_{2H}\right|^2+\left|\mathbf{J}_{3H}\right|^2-\left|u_{1X}\right|^2-\left|u_{2X}\right|^2-\left|\mathbf{J}_{1X}\right|^2-\left|\mathbf{J}_{2X}\right|^2\\
    \dot{u}_{2H}&=-2u_{2H}^2-2\mathbf{J}_{1H}\cdot\mathbf{J}_{2H}+2u_{3H}\sum_jJ_{3H}^{(j)}+2u_{2H}\sum_jJ_{1H}^{(j)}-2\text{Re}\left[u_{1X}^*u_{2X}+\mathbf{J}^*_{1X}\cdot\mathbf{J}_{2X}\right]\nonumber\\
    \dot{u}_{3H}&=4u_{1H}u_{3H}-2u_{2H}u_{3H}+2\mathbf{J}_{1H}\cdot\mathbf{J}_{3H}+2u_{3H}\sum_jJ_{1H}^{(j)}+2u_{2H}\sum_jJ_{3H}^{(j)}\nonumber\\
    \dot{J}^{(x)}_{1H}&=2\left(u_{3H}J^{(x)}_{3H}-u_{2H}J^{(x)}_{2H}+2J^{(y)}_{1H}J^{(z)}_{1H}+J^{(y)}_{2H}J^{(z)}_{2H}+J^{(y)}_{3H}J^{(z)}_{3H}\right)-2\text{Re}\left[u_{1X}^*J^{(x)}_{1X}+u_{2X}^*J^{(x)}_{2X}-J^{(y)*}_{1X}J^{(z)}_{1X}-J^{(y)*}_{2X}J^{(z)}_{2X}\right]\nonumber\\
    \dot{J}^{(x)}_{2H}&=2\left(-u_{2H}J^{(x)}_{1H}-J^{(x)2}_{2H}+J^{(y)}_{2H}J^{(z)}_{1H}+J^{(y)}_{1H}J^{(z)}_{2H}\right)+2J^{(x)}_{3H}\left(u_{3H}-J^{(y)}_{3H}-J^{(z)}_{3H}\right)+2J^{(x)}_{2H}\left(J^{(x)}_{1H}-J^{(y)}_{1H}-J^{(z)}_{1H}\right)-\nonumber\\
    &-2\text{Re}\left[u_{1X}^*J^{(x)}_{2X}+u_{2X}^*J^{(x)}_{1X}-J^{(y)*}_{1X}J^{(z)}_{2X}-J^{(y)*}_{2X}J^{(z)}_{1X}\right]\nonumber\\
    \dot{J}^{(x)}_{3H}&=2\left(2u_{1H}J^{(x)}_{3H}+u_{3H}J^{(x)}_{1H}+J^{(y)}_{1H}J^{(z)}_{3H}+J^{(y)}_{3H}J^{(z)}_{1H}-J^{(x)}_{2H}J^{(x)}_{3H}\right)+2J^{(x)}_{2H}\left(u_{3H}-J^{(y)}_{3H}-J^{(z)}_{3H}\right)+2J^{(x)}_{3H}\left(J^{(x)}_{1H}-J^{(y)}_{1H}-J^{(z)}_{1H}\right)\nonumber
\end{align}
The flow equations for \(J^{(y)}\) and \(J^{(z)}\) are obtained by cyclic permutation from the \(J^{(x)}\) equations. The coupling constants within the \(V\) chain have the same flow equations with \(H\) replaced by \(V\). Observe that even if at the bare level \(u=0\), they are generated by the \(J\)'s, and so have to be included in the analysis even if we are mostly interested in the spin fluctuations. Observe also that the flows for the umklapp couplings are not affected by the inter-chain interactions. The inter-chain coupling constants flow as follows:
\begin{align}
    \dot{u}_{1X}&=-u_{1v}u_{1X}-u_{2V}u_{2X}-u_{1X}^*u_{1H}-u_{2X}^*u_{2H}-\mathbf{J}_{1V}\cdot\mathbf{J}_{1X}-\mathbf{J}_{2V}\cdot\mathbf{J}_{2X}-\mathbf{J}^*_{1X}\cdot\mathbf{J}_{1H}-\mathbf{J}^*_{2X}\cdot\mathbf{J}_{2H}\nonumber\\
    \dot{u}_{2X}&=-u_{1V}u_{2X}-u_{2V}u_{1X}-u_{2X}^*u_{1H}-u_{1X}^*u_{2H}-\mathbf{J}_{1V}\cdot\mathbf{J}_{2X}-\mathbf{J}_{2V}\cdot\mathbf{J}_{1X}-\mathbf{J}^*_{2X}\cdot\mathbf{J}_{1H}-\mathbf{J}_{1X}^*\cdot\mathbf{J}_{2H}\nonumber\\
    \dot{J}^{(x)}_{1X}&=-u_{1V}J^{(x)}_{1X}-u_{1X}J^{(x)}_{1V}-u_{2V}J^{(x)}_{2X}-u_{2X}J^{(x)}_{2V}-u_{1X}^*J^{(x)}_{1H}-u_{1H}J^{(x)*}_{1X}-u_{2X}^*J^{(x)}_{2H}-u_{2H}J^{(x)*}_{2X}+\nonumber\\
    &+J^{(y)}_{1V}J^{(z)}_{1X}+J^{(y)}_{2V}J^{(z)}_{2X}+J^{(y)*}_{1X}J^{(z)}_{1H}+J^{(y)*}_{2X}J^{(z)}_{2H}+J^{(z)}_{1V}J^{(y)}_{1X}+J^{(z)}_{2V}J^{(y)}_{2X}+J^{(z)*}_{1X}J^{(y)}_{1H}+J^{(z)*}_{2X}J^{(y)}_{2H}\nonumber\\
    \dot{J}^{(x)}_{2X}&=-u_{1V}J^{(x)}_{2X}-u_{2X}J^{(x)}_{1V}-u_{2V}J^{(x)}_{1X}-u_{1X}J^{(x)}_{2V}-u_{2X}^*J^{(x)}_{1H}-u_{1H}J^{(x)*}_{2X}-u_{1X}^*J^{(x)}_{2H}-u_{2H}J^{(x)*}_{1X}+\nonumber\\
    &+J^{(y)}_{1V}J^{(z)}_{2X}+J^{(y)}_{2V}J^{(z)}_{1X}+J^{(y)*}_{2X}J^{(z)}_{1H}+J^{(y)*}_{1X}J^{(z)}_{2H}+J^{(z)}_{1V}J^{(y)}_{2X}+J^{(z)}_{2V}J^{(y)}_{1X}+J^{(z)*}_{2X}J^{(y)}_{1H}+J^{(z)*}_{1X}J^{(y)}_{2H}
\end{align}
Notice that the equations are clearly symmetric under the exchange of \(H\) and \(V\), implying that there may be fixed trajectories for which \(g_H\) and \(g_V\) are equal. Assuming isotropic spin-fluctuations then leads to Eqs. (\ref{RGflowEq}).
\end{widetext}

\bibliography{UTe2}

\end{document}